\documentclass[twocolumn,showpacs,preprintnumbers,amsmath,amssymb,fleqn]{revtex4}

\usepackage{graphicx}
\usepackage{dcolumn}
\usepackage{bm}

\begin{document}

\title{Anti-crossings of spin-split Landau levels in an InAs \\ two-dimensional electron gas with spin-orbit coupling}

\author{W. Desrat}
\author{F. Giazotto}
\author{V. Pellegrini}
\author{M. Governale}
\author{F. Beltram}
\affiliation{NEST-INFM and Scuola Normale Superiore, Piazza dei Cavalieri 7, I-56126 Pisa, Italy}

\author{F. Capotondi}
\altaffiliation[Also at ]{Dipartimento di Fisica, Universit\`a di Modena e Reggio Emilia, I-43100 Modena, Italy}

\author{G. Biasiol}
\author{L. Sorba}
\altaffiliation[Also at ]{Dipartimento di Fisica, Universit\`a di Modena e Reggio Emilia, I-43100 Modena, Italy}
\affiliation{NEST-INFM and Laboratorio Nazionale TASC-INFM, Area Science Park, I-34012 Trieste, Italy}

\date{\today}

\begin{abstract}
We report tilted-field transport measurements in the quantum-Hall regime in an InAs/In$_{0.75}$Ga$_{0.25}$As/In$_{0.75}$Al$_{0.25}$As quantum well.
We observe anti-crossings of spin-split Landau levels, which suggest a mixing
of spin states at Landau level coincidence. We propose that the level repulsion is due to the presence of
spin-orbit and of band-non-parabolicity terms which are relevant in narrow-gap systems. Furthermore, electron-electron interaction is significant in our structure, as demonstrated by the large values of the interaction-induced enhancement of the electronic $g$-factor.
\end{abstract}


\maketitle

A rich class of magnetic phenomena in two-dimensional electron systems (2DES)
due to the crossing of Landau levels with opposite spin polarizations has recently been the subject of intense
theoretical and experimental efforts. Giuliani and Quinn \cite{Giuliani85}
predicted a first-order transition between spin-polarized and spin-unpolarized quantum Hall (QH)
states near crossing between Landau levels in a tilted-field configuration at $\nu=2$.
By tilting the field, the ratio between the Zeeman energy ($E_z$) and the cyclotron energy ($E_c$)
increases, eventually reaching unity. It was shown, however, that the transition occurs before
coincidence of the single-particle energy levels. It happens when the energy
difference between the last occupied and first unoccupied single-particle Landau levels
reaches a critical value, which depends on the strength of electron-electron interaction \cite{Giuliani85,Koch93}.

It has been shown that the first-order transition always occurs when the two Landau levels involved in the crossing
have similar wavefunction profiles like in the single-layer single-subband systems \cite{Jungwirth98,Jungwirth00}.
In this case the gain in the exchange energy at the transition dominates over Hartree and single-particle
energy splitting, leading to an easy-axis anisotropy and to Ising ferromagnetism \cite{Jungwirth00}.
In the opposite limit, which can be reached, for instance,  in a double-layer system, a second-order phase transition
can occur, leading to an easy-plane anisotropic QH ferromagnetic state \cite{Jungwirth00}. In magneto-transport studies, Ising ferromagnetism has remarkable
experimental signatures. It is associated either to the complete disappearing of the QH state at the crossing or
to the persistence of the QH state with resistance spikes and hysteretic behavior \cite{Piazza99,DePoortere00}.
It is now established that these effects stem from the dynamics of magnetic domains, and are strongly affected
by disorder or temperature. Most of the experimental efforts focused on high-mobility GaAs/AlGaAs heterostructures, but also other materials were exploited in order to have larger $g$-factors, so that Zeeman and cyclotron energies
become comparable at more experimentally accessible magnetic fields. To this end, AlAs
quantum wells and II-VI diluted magnetic semiconductors were studied \cite{DePoortere00,Jaroszynski02}.
Transitions between spin-split Landau levels have also been reported in InAs-based single-layer
systems like in InGaAs/InP heterostructures \cite{Koch93} and in InAs/AlSb quantum wells \cite{Brosig00}.

The spin properties of QH ferromagnets could be radically modified by the presence of additional spin-mixing
terms. A few theoretical works investigated the impact
of single-particle effects related to Rashba and Dresselhaus spin-orbit couplings on the spin properties
of QH states at low filling factors \cite{Falko92,Falko93,Schliemann03}. In Ref.\cite{Falko93}, an anti-crossing of the two lowest spin-split Landau levels due to a new partially spin-polarized spin-density-wave-like state induced by spin-orbit coupling was predicted. These theoretical works indicate that
single-particle effects cooperate with Hartree and exchange
contributions in determining the spin configuration of QH states. The experimental investigation of the
interplay between many-body effects and spin-mixing terms has not  been carried out yet. For this purpose,
semiconductor heterostructures displaying both large exchange interaction and spin-orbit coupling should be realized.

Here we report tilted-field magneto-transport measurements in a
2DES confined in a narrow InAs/In$_{0.75}$Ga$_{0.25}$As single
quantum well which exhibits significant
spin-orbit coupling and large exchange interaction.
Anti-crossing of spin-split Landau levels at even and
odd low filling factors are observed close to degeneracy between
Landau levels with similar wavefunction profiles.
We argue that
these results offer evidence of the impact of spin-orbit
(SO) coupling on the collective spin
configuration of QH ferromagnets. The experiments also suggest
that additional single-particle spin-mixing terms, originating
from band non-parabolicity (NP)  (usually neglected at low
magnetic fields), can also determine the collective spin
configuration of QH states.

\begin{figure}[]
\includegraphics[width=8cm]{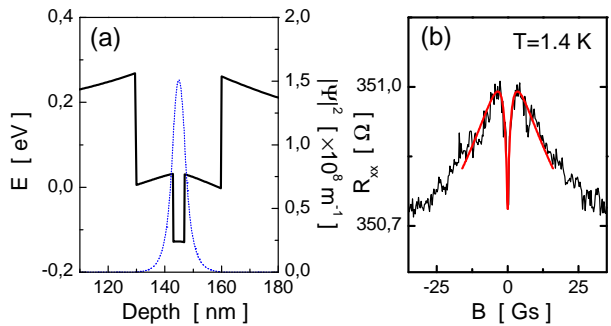}
\caption{\label{fig1} (a) Conduction band (left axis) and density profile (right axis) for the
InAs/In$_{0.75}$Ga$_{0.25}$As quantum well determined from 1D a Poisson-Schr\"odinger simulation
with a constant background doping $N_{dd}=2.9\times10^{16}$~cm$^{-3}$. (b) Experimental and
theroretical low-magnetic-field dependences of the longitudinal resistance (black and red curves
respectively) at $T=1.4$ K.}
\end{figure}

\begin{figure}[]
\includegraphics[width=8cm]{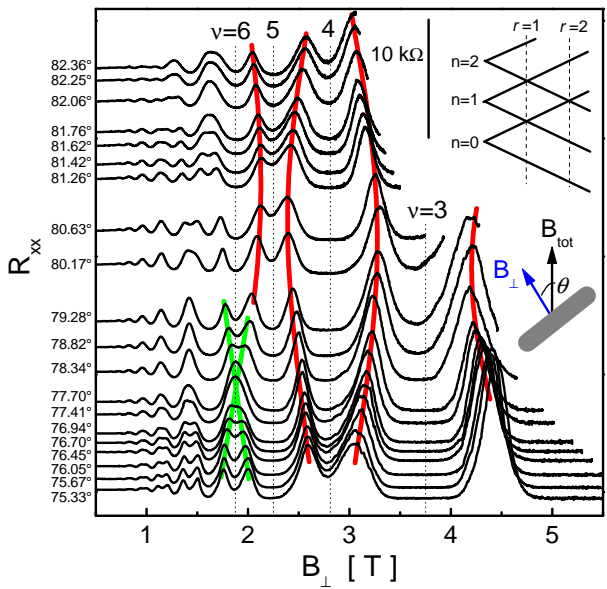}
\caption{\label{fig2} Longitudinal resistance as a function of the
perpendicular magnetic field for different tilt angles measured at
$T=30$ mK. Curves have been shifted proportionally to
$1/\cos(\theta)$ for clarity. Vertical thin dotted lines indicate
constant integer filling factors. Crossing at $\nu=6$ and
anti-crossings at $\nu=3$ and $5$ are enlightened by the green and
red eye-guide lines, respectively. Upper inset: schematic diagram of the
spin-split Landau levels vs $1/\cos(\theta)$ (i.e., vs $E_z$ for fixed perpendicular magnetic field).
Lower inset: sketch of the tilt-angle geometry.}
\end{figure}

The sample consists of an unintentionally doped 30-nm-wide
In$_{0.75}$Ga$_{0.25}$As quantum well embedded in
In$_{0.75}$Al$_{0.25}$As barriers and metamorphically grown on a
GaAs substrate \cite{Capotondi03}.
A thin 4 nm InAs layer is
inserted in the center of the well. The conduction band and
density profiles computed by a 1D Poisson-Schr\"odinger solver
\cite{Snider} are shown in Fig.\ref{fig1}(a).
The 2DES is
characterized by a carrier density of
$n_s=2.73\times10^{11}$~cm$^{-2}$ and a mobility of
$\mu=1.93\times10^5$ cm$^2$/Vs, which can be tuned by biasing a top gate.
Taking into account the
confinement correction and the weight of the
wavefunction in the different layers, a five-band
$\boldsymbol{k\cdot p}$ calculation yields $m^*=0.034\,m_0$
(where $m_0$ is the free electron mass) and $g^*=-10.4$ for
the effective mass and the $g$-factor, respectively.
Magneto-transport measurements in the QH regime were performed by placing the sample in
a dilution fridge ($T_{bath} =30$~mK) where it can be rotated
$in\, situ$ with respect to the magnetic field (the tilt-angle geometry is shown in the lower part of Fig.\ref{fig2}).
The longitudinal
resistance is measured on a 80-$\mu$m-wide Hall bar with phase-sensitive technique using a 20 nA bias current.
Figure \ref{fig2}
displays the longitudinal magnetoresistance $R_{xx}$ versus the perpendicular
magnetic field $B_\perp$ for different tilt
angles $\theta$ (the latter are accurately determined from the
Hall voltage).

In Fig.\ref{fig2} several
transitions between spin-split Landau levels are observed at even
and odd filling factors. These coincidences occur each time the
ratio between the exchange-enhanced Zeeman and cyclotron energies
is an integer $r=E_z/E_c=1,\,2,\, \ldots$, as shown schematically
in the right upper part of Fig.\ref{fig2}. First, we focus  on the transition
at $\nu=6$ corresponding to $r=1$. Two distinct peaks are observed
at very low tilt angles (lying at $B\approx 1.75$ and 2 T) that
merge into a single peak at the coincidence $\theta\sim
77.4^\circ$, and split again at larger angles (green
eye-guide lines). This scenario is also observed for higher filling
factors, $\nu=7$ and 8. Conversely, the evolutions at $\nu=3$
and 5 are characterized by anti-crossing as a function
of the tilt angle (red eye-guide lines in
Fig.\ref{fig2}).

\begin{figure}[]
\includegraphics[width=8cm]{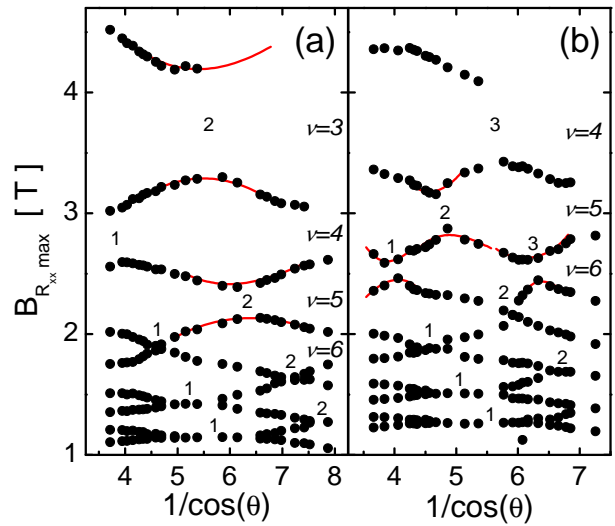}
\caption{\label{fig3} Perpendicular magnetic field position of the longitudinal resistance peaks as a function
of the inverse cosine of the tilt angle at gate voltages (a) $V_g=0$ V ($n_s=2.73\times10^{11}$~cm$^{-2}$)
and (b) $V_g=0.5$ V ($n_s=3.65\times10^{11}$~cm$^{-2}$). Thin red eye-guide lines have been added to emphasize the
anti-crossings. The crossings and anti-crossings have been labeled with the corresponding value of $r$.}
\end{figure}

Figure \ref{fig3}(a) shows the perpendicular magnetic field position
of the $R_{xx}$ peaks, $B_{R_{xx} max}$, as a function of
$1/\cos(\theta)$.
Crossings and anti-crossings are labeled with the corresponding $r$.
It is clear
that at $\nu=5$ two adjacent Landau levels  come closer
and then repel avoiding the crossing.
Even if the
measure of the gap at the coincidence is presently lacking,
we stress that at $\nu=5$ ($r=2$) the resistance minimum does not reach
zero at this very low temperature (see Fig.\ref{fig2}),
indicating that the gap is relatively small. A qualitative
comparison with the $R_{xx}(B)$ curves obtained on a similar sample where
 activation energies have been measured at several integer
filling factors (data not shown), suggests that the gap at the
transition should be less than 0.5 K.

In Fig.\ref{fig3}(b) the $R_{xx}$ peak position is shown with the top-gate
biased at $V_g=0.5$ V. This positive voltage slightly
increases the carrier density up to
$n_s=3.65\times10^{11}$~cm$^{-2}$ and shifts the filling factors
toward higher perpendicular fields. As for the zero bias case,
the large filling factors ($\nu=7,\,8,\,9$ and 10) exhibit
conventional crossings whereas the low fillings ($\nu=4$ and $5$)
present anti-crossings. We note that the transition at $\nu=6$
appears now as an anti-crossing.
The second transition observed at $\nu=6$ at higher tilt angle
corresponds to $r=3$. Thus the effective Zeeman energy is 3 times
larger than the cyclotron energy implying that the electronic
$g$-factor is large in our system (e.g., $\vert
g^*\vert=3\times(2\cos(\theta)/m^*)=28\pm0.5$) as expected from
the presence of the narrow-gap InAs layer. Furthermore, the
strong filling factor dependence of the coincidence conditions at
fixed $E_z/E_c$ ratio (e.g. at $r=1$,
$1/\cos(\theta_c)=3.97,\,4.56$ and 5.07 for $\nu=6,\,8$ and 10
respectively, where $\theta_c$ is the coincidence angle) highlights the large contribution of the
electron-electron interaction to the spin gap, as already estimated
elsewhere \cite{Desrat04}.

Our system exhibits a finite spin-orbit coupling
as expected for InGaAs heterostructures \cite{Zawadzki04}. Figure \ref{fig1}(b) shows the longitudinal
resistance as a function of the magnetic field at $T=1.4$ K and $\theta=0$ in the low field region. The resistance
enhancement upon reducing $B$ is due to weak localization, while the narrow dip around $B=0$ is due to
 weak anti-localization  induced by spin-orbit coupling \cite{Koga02,Miller03}. Figure \ref{fig1}(b) also reports the result of a theoretical calculation
based on the Iordanskii, Lyanda-Geller and Pikus  model \cite{ilp94}. From fitting the data, we estimate the
effective spin-orbit and phase-breaking magnetic fields, $H_{\text{so}}=1.5$ Gauss, $H_{\text{so}}^\prime=0.8$ Gauss
 and $H_{\phi}=0.067$ Gauss. Figure \ref{fig1}(b) clearly demonstrates that
 spin-orbit terms are relevant in our system and co-exists with many-body interactions \cite{noteWAL}.

We are now in the position to discuss the possible origin of the
observed anti-crossings. To this end we recall that in a
single-layer system when two Landau levels come into coincidence
the exchange energy leads to an easy-axis ferromagnet
characterized by a non-vanishing gap at the transition. This gap
is proportional to the Coulomb interaction
$e^2/4\pi\varepsilon\ell_B$ ($\ell_B$ is the magnetic length and
$\varepsilon$ the dielectric constant) and is typically much
larger than $k_BT$ at millikelvin temperatures. Thus the common
signature of an easy-axis ferromagnet is the observation of a
persistent $R_{xx}=0$ region in the longitudinal magnetoresistance
\cite{Koch93}. Disorder, however, may lead to the formation of
partially and fully polarized magnetic domains separated by domain
walls. In the case of large wall loops, electrons, at coincidence, can diffuse along
the walls  and can thus backscatter from one edge of
the Hall bar to the other, giving rise to narrow resistance spikes in
the QH minima regions or even to complete breakdown of the QH
state \cite{Jungwirth01,Piazza99,DePoortere00,Muraki01}. Spikes
and hysteresis of the resistance value with relaxation in time are
often associated to the slow and complex domain wall motion
\cite{Piazza03,DePoortere03}. Assuming that electron-electron
interaction dominates, one may speculate that the anti-crossing-like
behavior observed at low filling factors is consistent with the
formation of an easy-axis ferromagnet at the transition. This
would be rather surprising since the gap associated to an
easy-axis ferromagnet is expected to be a fraction of the Coulomb
energy which is as large as 70 K at $B=2.5$ T, i.e., much larger
than the experimental temperature. The fact that the longitudinal
resistance does not reach zero in spite of this large gap could
arise from disorder-induced broadening of the Landau levels which in the present system
is bigger than in usual high-mobility samples. However, the lack
of direct manifestations of easy-axis ferromagnetism, such as
resistance spikes or hysteresis in Shubnikov-de Haas traces, is not
consistent with this scenario.

In the following we argue that anti-crossing can instead arise from spin-mixing terms, eventually leading
to easy-plane ferromagnetism \cite{Jungwirth98,Jungwirth00,Muraki01}. Before discussing the impact of these
terms it is important to notice that easy-plane ferromagnetism can also occur in the absence of
spin-mixing terms when a sufficiently large Hartree energy characterizes the transition.
This is however improbable in  our system, given the very similar wavefunction profiles of the two
degenerate Landau levels along the direction perpendicular to the 2DES \cite{Jungwirth98,Jungwirth00}.
We note that the appearance of an easy-plane anisotropy in a single-layer system may occur at very large
tilt angles only when the orbital effect of the in-plane magnetic field leads to a modification of the
density profile \cite{Jungwirth98}. In Fig.\ref{fig3}(b), the first anti-crossing at $\nu=6$ is observed at
$\theta_c\approx75.4^\circ$ for which the impact of the in-plane field is still weak.
To describe the
repulsion of Landau levels close to the coincidence additional effects
need to be considered. SO coupling can play an important role \cite{Falko92,Falko93,Schliemann03}
but it cannot couple any pair of Landau levels \cite{Das90,Chokomakoua04}. Whereas the levels mixing at
$r=1$ and $r=3$ (i.e., the difference between the orbital numbers of the degenerate levels is $\Delta n =1$
and $\Delta n =3$ respectively) is allowed and can lead to anti-crossings at even filling factors, the coupling
of Landau levels with $\Delta n=2$, $r=2$ is prohibited. Therefore, in order to describe the anti-crossing observed
at $\nu=5$ we need to consider additional terms with the appropriate selection rules. For example,
terms arising from band non-parabolicity (NP), such as $g'' \mu_B \frac{1}{\hbar^2} \{\boldsymbol{\sigma \cdot \Pi},
\boldsymbol{B \cdot \Pi}\}$, where $\boldsymbol{\sigma}$ is the vector of the Pauli matrices,
$\boldsymbol{\Pi}=\boldsymbol{p}+q\boldsymbol{A}$ the canonical momentum and the curly brackets stands for
the anti-commutator \cite{Kim89}.
Selection rules and anti-crossing-energy gaps
can be obtained considering the three-dimensional NP [see, for example,
Eq. (5) of Ref. \cite{Kim89}] and Dresselhaus terms, projecting them on the two-dimensional plane,
and computing matrix elements of the resulting Hamiltonian between
Landau levels after the inclusion of the Rashba contribution.
Using typical $k\cdot p$ estimates for the NP coupling constants and considering the typical magnetic fields in our experiment, anti-crossing gaps $\sim0.5$ K can be obtained consistent with the experimental findings at $\nu =5$ ($r=2$). The values of the anti-crossing gaps derived from SO coupling can be estimated from the weak anti-localization measurements.
We found values of the order of 1 K consistent with the observed anti-crossing behavior at $r=1$ and $r=3$.

For a more quantitative comparison with the
experimental results shown in Figs.\ref{fig2} and \ref{fig3}, a
many-body approach along the lines of
Refs.\cite{Jungwirth98}, \cite{Falko93} needs to be carried out.
Such a task is out of the scope of the present work. The
single-particle model however highlights that transitions at $r=1$
are governed by the $k$-linear terms (Rashba and linear Dresselhaus
terms \cite{Das90,Kim89}) which couple adjacent Landau levels.
Similarly, the model suggests that the anti-crossings
at $r=2$ are due to NP terms, while those
at $r=3$ to the cubic terms of the Dresselhaus interaction. The NP
terms are usually
neglected in experiments on spin-mixing effects in semiconductors.
They, however, play a significant role in the present experiment due to the
large magnetic fields.

Finally we point out that the proposed scenario relating level repulsion with spin-mixing terms is also consistent with the tilted-field results obtained in InGaAs heterostructures with no InAs layers (lower SO coupling)  where no anti-crossings were observed \cite{Desrat04}.

In conclusion, longitudinal magnetoresistance traces in a tilted magnetic field
configuration on an InAs/InGaAs quantum well exhibit anti-crossings of spin-split
Landau levels at low filling factors. We propose that spin-orbit
interaction and non-parabolicity effects can induce the observed anti-crossings. The microscopic nature of the
quantum Hall ground states originating from the interplay between exchange energy
and spin-mixing terms remains an open intriguing issue for further investigations.

We are indebted to D.K. Maude for his help at the Grenoble High Magnetic Field Laboratory
where the tilted field measurements have been carried out. We acknowledge Rosario Fazio, Diego Frustaglia
and Giuseppe La Rocca for fruitful discussions. This work was supported in part by the
Ministry of University and Research (MIUR) under the FIRB "Nanotechnologies and nanodevices
for the information society" and COFIN programs and by the European Research and Training
Network COLLECT (Project HPRN-CT-2002-00291).

\bibliography{bibv2}

\begin{thebibliography}{26}
\bibitem[{not()}]{noteWAL}
\bibinfo{note}{A quantitative analysis of Dresselhaus ans Rashba spin-orbit
  coupling constants from weak anti-localization measurements will be reported
  elsewhere.}
\end{thebibliography}

\end{document}